\documentstyle[twocolumn,psfig,rotating]{mn}

\newif\ifAMStwofonts
\AMStwofontstrue

\def\phai{{\it Phase I}}
\def\phaii{{\it Phase II}}
\def\phaiii{{\it Phase III}}
\def\kms{\rm{km s^{-1}}}

\def\caH{{\cal H}}

\def\cwsh{ C_{_{\rm wsh}}}
\def\cd{ C_{_{\rm d}}}

\def\caSi{{\cal S}_{_{\rm I}}}

\def\gsim{~\rlap{$>$}{\lower 1.0ex\hbox{$\sim$}}}

\def\rv{{R_{\rm v}}}

\def\ltsim{\lower.5ex\hbox{$\; \buildrel < \over \sim \;$}}
\def\gtsim{\lower.5ex\hbox{$\; \buildrel > \over \sim \;$}}
\def\ltsim{\lower.5ex\hbox{$\; \buildrel < \over \sim \;$}}
\def\gtsim{\lower.5ex\hbox{$\; \buildrel > \over \sim \;$}}

\def\rej{{   R_{_{\rm inj}}  }}
\def\dnej{{   {\dot N}_{_{\rm inj}}  }}
\def\rhoi{{\rho_{_{\rm I}}}}
\def\rhob{{\rho_{_{\rm b}}}}
\def\rhoa{{\rho_{_{\rm a}}}}
\def\gammai{{\gamma_{_{\rm I}}}}
\def\gammaa{{\gamma_{_{\rm a}}}}
\def\gammab{{{\gamma_{_{\rm b}}}}}
\def\ui{{u_{_{\rm I}}}}
\def\ub{{u_{_{\rm b}}}}
\def\ua{{u_{_{\rm a}}}}
\def\uA{{u_{_{\rm A}}}}

\def\kms{\mbox{km\,s$^{-1}$}}

\def\dd{\,{\rm d}}

\def\rb{{R_{_{\rm b}}}}
\def\gammab{\gamma_{_{\rm b}}}
\def\vb{{v_{_{\rm b}}}}
\def\nb{{n_{_{\rm b}}}}

\def\bnb{{{\bar n}_{_{\rm b}}}}
\def\pb{{p_{_{\rm b}}}}

\def\eb{{E_{_{\rm b}}}}
\def\ebi{{E_{_{\rm bi}}}}
\def\pbi{{p_{_{\rm bi}}}}
\def\rbi{{R_{_{\rm bi}}}}

\def\rhob{{\rho_{_{\rm b}}}}

\newcommand{\etal}{{\it et. al.}\ }

\include{rotating}

\begin{document}
\textheight 9in
\title[Suppressing cluster cooling flows]{Suppressing cluster cooling flows by self-regulated heating from spatially distributed population of AGNs}

\author[Nusser, Silk \& Babul] { Adi Nusser$^{1}$, Joseph Silk$^{2}$ 
and Arif Babul$^3$ \\\\
  $^{1}$Physics Department- Technion, Haifa 32000, Israel\\
  $^{2}$Astrophysics, Oxford University, Keble Road, Oxford OX1 3HR, UK\\
  $^3$ Department of Physics and Astronomy, University of Victoria, Victoria, BC V8P 5C2, Canada}

\maketitle

\begin{abstract}
Existing models invoking AGN activty to resolve the cooling flow conundrum in galaxy clusters 
focus exclusively on the role of the central galaxy.    Such models require fine-tuning of 
highly uncertain microscopic transport properties to distribute the thermal thermal over the 
entire cluster cooling core.   We propose that the ICM is instead heated by multiple, spatially 
distributed AGNs.  The central regions of galaxy clusters are rich in spheroidal systems,
all of which are thought to host black holes and could participate in the heating of the 
ICM via AGN activity of varying strenghts.  And they do.  There is mounting observational
evidence for multiple AGNs in cluster environments.
Active AGNs drive bubbles into the ICM.  
We identify three distinct interactions between the bubble and the ICM: (1) Upon injection, the 
bubbles expand rapidly in situ to reach pressure equilibrium with their surroundings, generating 
shocks and waves whose dissipation is the principal source of ICM heating.  (2) Once inflated, the 
bubbles rise buoyantly at rate determined by a balance with the viscous drag force, which itself 
results in some additional heating. (3) Rising bubbles expand and compress their surroundings.   
This process is adiabatic and does not contribute to any additional heating;  rather, the increased 
ICM density due to compression enhances cooling.
Our model sidesteps the ``transport'' issue by relying on the spatially distributed galaxies 
to heat the cluster core.  We include self-regulation in our model by linking AGN activity in a 
galaxy to cooling characteristics of the surrounding ICM.  We use a spherically symmetric 
one-dimensional hydrodynamical code to carry out a preliminary study illustrating the efficacy of 
the model.  Our self-regulating  scenario predicts that there should be enhanced AGN activity of 
galaxies inside the cooling regions compared to galaxies in the outer parts of the cluster.
This prediction remains to be confirmed or refuted by observations.

\end{abstract}
 
                           
\section{introduction}
The observed properties of the x-ray emitting gas in clusters of galaxies
present several challenging puzzles for which satisfactory explanations are 
yet to be found. The absence of cold gas (with temperatures below $\sim 1$ keV) is 
one of those puzzles (e.g. Peterson \etal 2001).
Because of radiative losses, one would expect  
the  temperature profiles to fall below
 a keV or so in the central regions where cooling
is most efficient. Yet in none of the observed clusters does the temperature  drop to the
level dictated by cooling alone. Hence   efficient heating mechanisms
must operate without fail at  the cores of all cooling clusters.  

The most popular mechanism for suppressing cooling is energy released by an  AGN harboured 
by the central cluster galaxy (cf.  Quilis \etal 2001, Babul \etal 2002, Kaiser \& Binney 
2003, Dalla Vecchia \etal 2004, Omma \etal 200, 
Roychowdhury \etal 2004, Ruszkowski \etal 2004,  Voit \& 
Donahue 2005, Br\"uggen \etal 2005, Nipoti \& Binney 2005, Roychowdhury \etal 2005, 
Sijacki \& Springel 2006, and references therein).  Relativistic ejecta from the AGN 
transform into hot bubbles that eventually reach pressure equilibrium with the ICM and 
proceed to rise buoyantly away from the center. These bubbles are believed to be one of 
the means with which the AGN can heat the ICM.  For example,  mechanical activity associated
with the inflation of the bubbles  near the center can generate internal gravity waves 
(ie.~disturbances that are subject to a restoring force of gravity or buoyancy)
(Omma \etal 2004) and/or sound waves which are believed to eventually dissipate their energy 
in the ICM. To balance cooling in a cluster of   x-ray luminosity of $L_{\rm x}\sim 
10^{44}\rm erg \; s^{-1}$, a central AGN must produce $\sim 10^{60} \rm erg$ 
over the entire life-time of the cluster. This is  at the upper limit of  the observed 
range of AGN energy output in galaxy clusters, based on the $pV$ content of x-ray cavities 
(e.g. B\^irzan \etal 2004). Since the $pV$ energy content must be a lower estimate of the 
AGN energy output (see below), it is reasonable to assume that cooling is quenched by AGN 
feedback.  The challenge, however, is to arrange for  efficient energy transport
from the AGN over the entire cooling core, or  out to distances of up to $\sim 100 \rm kpc$. 

Typically, studies investigating the impact of AGNs have tended to focus on AGN activity in 
only the central galaxy, and invoke relatively high values of thermal conduction to redistribute 
the energy across the cooling core (see, for example, Roychowdhury \etal 2005).  Here, we propose 
instead that the cluster ICM is heated by the AGN activity in more than just the one central
galaxy.  The central region of the clusters is rich in giant elliptical galaxies,  most of which 
are presumed to harbour supermassive black holes that are thought to inject radio bubbles into
the ICM  with a duty cycle of $\sim 10^8$ yr per Hubble time (Best \etal 2006).  

Upon injection, the overpressurized
bubbles rapidly expand (virtually in situ) to reach pressure equilibrium with their surroundings, 
generating waves and shocks, the dissipation of which is the primary mechanism by which the
ICM is heated.  Once in pressure equilibrium, the bubbles will rise buoyantly.  The dissipation of
the work done by viscous drag acting to retard the bubbles' motion will further heat the ICM, though 
to a much lesser extent.  In response to the decreasing pressure,  the rising bubbles will expand.  This expansion 
causes a corresponding compression of the ICM.  However, while the initial inflation of the cavity 
may trigger internal gravity waves, compression waves and shocks that heat the ICM, this subsequent 
expansion-compression is adiabatic in character and no heating of the ICM ensues, even though the 
ICM's internal energy will increase due to compression.  In fact, the compression actually 
leads to  the erosion of the thermal content of the ICM because the resulting increase in 
density results in enhanced cooling.
This can be seen as follows. The compression of the 
ICM can be approximated as an adiabatic process so that the ``entropy" $S=T/n^{2/3}$ is conserved.
The bremsstrahlung cooling time is $t_{\rm c}\propto T/(n T^{0.6})\sim 
S ^{0.4}n^{-1.06}$. Therefore, cooling becomes more efficient as 
$n$ is increased during  adiabatic compression. 

In our model, the heating occurs near where the bubbles are produced but
since the galaxies that produce these bubbles are distributed throughout
the cluster core region, so too is the heating.   Modest heat conduction
should suffice to distribute further the energy over the ICM between the
galaxies.  Since the energy deposition is a local phenonemon, 
it is relatively straightforward to construct a self-regulating model.

To model the interaction of the bubbles with the ICM (which we treat as 
all material not belonging to the hot and dilute bubbles), we use a spherical 
one-dimensional (1D) hydrodynamical code.  The code
treats the bubbles and the ICM as a two-fluid system 
in  a fixed dark matter halo. The code treats the interaction of  the bubbles and 
ICM in a consistent semi-analytical  way and assumes that the distribution of 
bubbles is spherically symmetric.
The code integrates the hydrodynamical equations of an {\it ambient medium}
with thermodynamic properties which depend on the local filling factor 
of bubbles. 
 
The outline of the paper is as follows. In \S\ref{sec:eom} we 
derive  the equations of motion of the ambient medium and present details of the 
the interaction of the bubble fluid with the ICM. In \S\ref{sec:self} we 
present our scheme for self-regulating feedback and 
discuss its  general properties. We describe the numerical implementation in
 \S\ref{sec:numerical} and  show results in  
\S\ref{sec:results}. We conclude with a summary and  general discussion of alternative heating schemes
in \S\ref{sec:sum}.
  
\section{The  equations of motion for the ambient medium}
\label{sec:eom}

A  1D hydrodynamical code cannot tackle the intrinsically three-dimensional (3D) 
problem of the  evolution of individual bubbles
in the ICM, in particular if bubbles are injected from sources away 
from the center as in our proposed scenario. 
Therefore, a formalism for following the evolution of bubbles statistically
must be developed. 
We represent bubbles as  a fluid specified by the 
following physical quantities: the radius of bubbles, their number density,
and velocity, all as a function of distance from the center.
The bubbles and the ICM are a two-fluid system that is best described
as a single fluid which we term the {\it ambient medium}.
The    representation in terms of  a single medium 
is applicable only  if local pressure equilibrium
between the bubbles and the ICM is established (Appendix A). 
Hereafter, quantities related to the ambient medium, the bubbles, and the ICM (gas 
outside bubbles) are denoted by the subscripts, $a$, $b$, and $I$, respectively. 
Suppose that the volume filling factor of bubbles at $r$ is $F(r)$.
The  ambient density, $\rhoa$, at position $r$,  is defined as
\begin{equation}
\rhoa=(1-F)\rhoi+F\rhob \; ,
\label{eq:rhoadef}
\end{equation}
in terms of the density inside bubbles, $\rhob$, and the density of the ICM, $\rhoi$, also at $r$.
The local ambient energy, $\ua$, per unit mass as
\begin{equation}
\ua=\frac{(1-F)\rhoi\ui+F\rhob\ub}{\rhoa} \; ,
\label{eq:uadef}
\end{equation}
where $\ub$ and $\ua$ are the energy per unit mass of the material inside the bubbles 
and of the ICM.
If the pressure, $p$, is 
$p=(\gamma-1)u\rho$ for both the ICM and the material inside the bubbles,
Following an analysis similar to that presented in  Appendix A  we find that the equation of state of the ambient 
medium is, 
\begin{equation}
p=(\gammaa-1)\rhoa\ua
\end{equation}
where 
\begin{equation}
\gammaa-1=\frac{(\gammai-1)(\gammab-1)}{(1-F)(\gammab-1)+F(\gammai-1)}\; ,
\end{equation} 
and we have assumed local pressure equilibrium  between the bubbles and the ICM.
These thermodynamic relations supplement the 
following equations of motion of the ambient medium in a gravitational field $g(r)$,
\begin{equation}
\frac{\dd v_{_{\rm a}} }{\dd t }= g -\frac{1}{\rhoa}\frac{\dd  p}{\dd r}\; .
\label{eq:euler}
\end{equation}
 In the absence of dissipative heating and cooling, the adiabatic energy equation is,
\begin{equation}
\frac{\dd \ua}{\dd t}=\frac{p}{\rhoa^{2}}\frac{\dd \rhoa}{\dd t} \; .
\label{eq:adiab}
\end{equation}
Dissipative heating and cooling will affect the energy equation as described below.
In the above equations the change of the filling factor as a result of 
dissipative heating and cooling, production of new bubbles, and bubble motion 
will have to be taken into account self-consistently.

\subsection{Motion of bubbles and their interaction with the ICM} 
\label{sec:drag}
In our model, cluster galaxies affect the ICM by the 
production of over-pressurised bubbles made of hot relativistic plasma with
an adiabatic index  $\gammab=4/3$. Bubbles are produced in response to local
cooling in the vicinity of galaxies according to the self-regulating mechanism 
described in \S\ref{sec:self}.

Let $\nb(r)$, $\vb(r)$, and $\rb(r)$, be, respectively, the number density of bubbles, the velocity of 
bubbles relative to the ambient medium,  and the radius of bubbles, all at distance $r$ from the center.
The motion of the bubbles is modeled as that of a pressure-free (i.e. collisionless) fluid with a velocity that is  determined
by a balance between buoyancy and drag forces with the ICM. 
Bubbles are assumed to be injected with the same energy per bubble  and
the same initial pressure  so that at distance $r$ all bubbles have the same 
size (see equation \ref{ref:radius}) and the same velocity.  Therefore, 
a hydrodynamical description for the flow of bubbles is 
self-consistent.  The pressure-free assumption is also self-consistent 
since $\vb(r)$ is a single-valued function of $r$ so that no "shocks" can form. 

The various phases of the evolution of the bubbles are as follows:
\begin{itemize}
\item {\phai:}
A bubble is injected at distance $r$ from the center.  This over-pressurised  bubble 
undergoes a  super-sonic expansion in the ICM until its internal pressure drops close 
to that of the ICM. During this phase of rapid expansion, the bubbles heat the ICM by 
the production of weak shocks.  This expansion is assumed to be quite rapid and such 
that the bubble does not move significantly away from the distance at which it has 
been injected.  
\item {\phaii:} The dilute hot bubble then rises by buoyancy towards 
more distant regions of lower pressure.  The moving bubbles heat the ICM via 
viscous drag forces.  As noted, the velocity of  bubbles relative to the ambient medium 
is determined by a balance between buoyancy and drag forces with the ICM. Rising bubbles
also expand adibatically, compressing the surrounding ICM.  This process does not heat 
the ICM\footnote{Deviation from adiabatic compression may result in 
bulk motions in the ICM. These motions can eventually dissipate into heat via turbulence or ordinary viscosity. The maximum heating rate 
that is obtained this way is slightly less that the drag heating rate discussed in \S\ref{sec:drag}}.   Increased ICM density does, however, elevate the efficiency of cooling.
\item {\phaiii:}
Finally, bubbles are destroyed by hydrodynamical (Raleigh-Taylor and Kelvin-Helmholtz) 
instabilities (e.g. Soker, Blanton \& Sarazin 2002).  
\end{itemize}
  
\subsubsection{Initial parameters of an injected bubble}

Let a bubble be  injected with 
an initial energy $\ebi$ and an initial internal pressure 
$\pbi$. 
Since $\ebi=\ub_i\rhob_i 4\pi\rbi^3/3$, where $\rhob_i$ and 
 $\ub_i$ are the initial mass density and
 energy per unit mass in the bubble, the equation of state $\pb=(\gammab-1)\rhob\ub$ yields,  
 \begin{equation}
 \rbi=\left[\frac{3}{4\pi}(\gammab-1)\frac{\ebi}{\pbi}\right]^{1/3} \; .
\label{ref:radius}
\end{equation}

\subsubsection{Bubble radius as a function of distance}
The bubble radius, $\rb(r)$, at any distance $r$ from the center can be estimated
from the adiabatic condition $\pb\rhob^{-\gammab}=const$. This condition  must hold 
in \phai\ and \phaii\ because the speed of sound of the 
relativistic plasma in the bubble is
much larger than its expansion velocity.
The adiabatic condition yields 
\begin{equation}
\rb(r)=\rbi\left[\frac{\pbi}{p(r)}\right]^{\frac{1}{3\gammab}} \; ,
\label{eq:rb}
\end{equation}
which for $\gammab=4/3$ gives the weak dependence, $\rb \propto (\pbi/p)^{1/4}$, on the pressure ratio.
 
\subsubsection{Heating by shocks}
\label{sec:shocks}

We  estimate the energy transferred from bubbles  to the ICM 
by means of weak shocks during \phai .
The equation of state of the material inside the bubble  
gives  $\eb=(4\pi/3){\rb}^3 \pb/(\gammab-1)$ for the thermal energy of a bubble
of radius $\rb$ with internal pressure $\pb$.
At the end of the rapid expansion in \phai\ we assume that $\pb\approx p(r)$
where $p(r)$ is the ICM pressure at the distance at which the bubble has been injected.
Substituting  $p=p(r)$ in (\ref{eq:rb}) to get $\rb$, we find that the thermal energy at the end 
of \phai\ is
\begin{equation}
\eb_0=\ebi \left(\frac{p}{\pbi}\right)^{1-1/\gammab}\; .
\label{eq:eb0}
\end{equation}
This equation implies that $\eb_0\propto p \rb^3$ can be substantially 
smaller than the initial energy of the bubble. 
The heat transferred to the ICM by the generation of shocks is
 \begin{equation}
 \Delta E_{\rm 1}=\cwsh(\ebi-\eb_0)=\cwsh 
 \ebi\left[1-\left(\frac{p}{\pbi}\right)^{1-\frac{1}{\gammab}}\right] \; .
\label{eq:Edissp}
\end{equation}
where $0\le \cwsh\le 1$ is an efficiency parameter. In the present treatment, we assume
that the energy transferred to the ICM is deposited locally.
The expression (\ref{eq:Edissp}) does not take into account the 
energy needed to place the bubble with  initial radius $\rbi $ and energy $\ebi$ into the ambient medium. For a given $\ebi$, this energy is negligible
if $\rbi$ is sufficiently small.

\subsubsection{Buoyancy and heating of the ICM by drag with rising bubbles} 
\label{sec:drag}
%%%%%%%%%%%%%%%%

The dynamical evolution of the system comprised of an expanding, rising bubble in a bubbly medium subject to radiative cooling is,
in general, a highly non-trivial problem, involving a host of complex phenomena including the compressible nature of the bubbly medium,
the evolution and dissipation of shocks, wakes and waves in this medium, bubble-bubble interactions mediated by bubble
wakes, bubble coalescence, changing bubble shapes and orientations, etc.  We write the equations of motion  of a bubble in \phaii .
We restrict ourselves to a simple situation of a system involving a single spherical bubble rising along the radial direction  in the ICM.
The equation of motion governing the bubble's rise is
\begin{equation}
\frac{d}{dt}\left(M_o\vb \right) = F_{_{\rm bu}} - F_{_{\rm drag}} \;
\label{eq:EoM}
\end{equation}
where the two terms on the right represent the buoyancy force acting on the bubble of radius $\rb$
and the drag force on the bubble as it moves through the ICM.  The left-hand term is the rate of
change of the Kelvin impulse (also known as the added-mass force) corresponding to the displacement of the ambient medium
due to the bubble's rising motion. Analytic expressions for the "virtual added-mass" of the bubble can be obtained for simplified motions. For a laminar flow generated by a spherical bubble moving in an incompressible fluid 
 the expression for the virtual added mass is (e.g. Kendoush 2003),
\begin{equation}
M_o=C_{AM} \frac{4\pi}{3}\rb^3 \rhoa \; , \;\;\; C_{AM}=\left(\frac{1}{2} + 3\frac{{\dot R}^2}{\vb^2}\right) \; ,
\end{equation}
where $\dot R$ is the rate of expansion of the bubble.  We have neglected the inertia of the bubble as it is much smaller
than the added-mass term.  

A bubble of radius $\rb$ at distance $r$ experiences a buoyant force of
\begin{equation}
F_{_{\rm bu}}=\frac{4\pi}{3}\rb^{3} \rhoa g_{_{\rm eff}} \; ,
\label{eq:fbu}
\end{equation}
where we have assumed that $\rhob\ll \rhoa$.  Here,
$g_{_{\rm eff}}$ is the gravitational force field (per unit mass) as measured in
a frame of reference attached to the ambient medium at distance
$r$.  The actual gravitational field,  $g$,
differs from $g_{_{\rm eff}}$ only when the system deviates  from hydrostatic equilibrium, such as in a cooling runaway where $g_{_{\rm eff}}
\approx 0$.   
We write the drag term on the bubble as it moves through the ICM with velocity
$\vb$ as
\begin{equation}
F_{_{\rm drag}}=\frac{\pi}{2}\cd \rb^{2}\rhoa \vb^{2}\;  .
\end{equation}
where $\cd$ is the drag coefficient and we have assumed that
$\vb\ll c_{s}$,
In our model
all bubbles at the same distance share the same velocity.
But  in a realistic situation, bubbles are generated with different
initial energies and at random directions.
So a bubble may collide with others. Hence, to model this effect we write   $\rhoa$  instead of $\rhoi$
in the drag equation.

Typically, the magnitude of  added-mass term is much smaller than the buoyancy and
the drag terms and therefore, to first order, the rise velocity of the bubbles can
be estimated by equating the drag and buoyancy forces (e.g. Churazov \etal 2001),
yielding:
\begin{equation}
\vb^{2}=\frac{8 g_{_{\rm eff}} \rb}{3\cd} \; .
\label{eq:vbub}
\end{equation}
The bubble radius, $\rb(r)$, as given by (\ref{eq:rb}) depends only on $p(r)$.
Therefore,   all bubbles with the same initial radius and internal pressure
share the same velocity at $r$, independently of the distance
at which they have been injected into the ICM.
When $\vb$ is non-negligible compared to the speed of sound in the
ambient medium, the drag becomes more efficient. Here we account for the
increase in the drag coefficient by using the following
expression for $\vb$
\begin{equation}
\vb^{2}=\frac{8 g_{_{\rm eff}} \rb}{3\cd} \left(1+
\frac{8 g_{_{\rm eff}} \rb}{3\cd c_{\rm s}^2}\right)^{-1}\; .
\label{eq:vb}
\end{equation}

As a bubble rises, the gravitational potential energy of the system (the bubble and
the surrounding medium) decreases.  A Rising bubble also expands and its internal energy decreases.  The total energy available as the bubble rises from position 1 (point of injection) to position 2 is
\begin{equation}
\Delta E_{{\rm 2}} = (E_{{\rm b1}} - E_{{\rm b2}}) + \int^{2}_{1} \frac{4\pi}{3}\rb^{3} \rhoa g_{_{\rm eff}} \, dr\; .
\end{equation}
where the first difference pair, corresponding to the change in the bubble's internal energy, it can be evaluated using
equation the condition $p/\rhob^\gammab=const$, for adiabatic changes of the bubble fluid as
\begin{equation}
(E_{{\rm b1}} - E_{{\rm b2}}) =
\ebi \left(\frac{p_{{\rm 1}}}{\pbi}\right)^{1-1/\gammab} \left[ 1 - \left(\frac{p_{{\rm 2}}}{p_{{\rm 1}}} \right)^{1-1/\gammab} \right] \; ,
\end{equation}
Assuming that the ambient medium is in hydrostatic equilibrium, the second term too can be
straightfowardly evaluated, yielding:
\begin{equation}
\int^{2}_{1} \frac{4\pi}{3}\rb^{3} \rhoa \frac{\partial\phi}{\partial r}\, dr =
\gammab (E_{{\rm b1}} - E_{{\rm b2}})\; ,
\end{equation}
where $\phi$ is the gravitational potential.
Consequently, $\Delta E_{{\rm 2}} = (1+\gammab) (E_{{\rm b1}} - E_{{\rm b2}})$.
This is the energy available to lift the
bubble against the drag force, overcome the inertia of displacing the surrounding fluid, as well as compress the ambient
medium.   

In general, it is not possible to determine how much of this energy that will go towards compressing the ambient medium and how much will towards heating the ICM, be it due to drag or through the eventual dissipation of the wakes and motions associated with the displacement of the ICM, unless one knows the details of the rise-expansion process.    In the limit that the bubble rises slowly, maintaining pressure equilibrium with its surroundings at all times, the bubble expansion/ICM compression is adiabatic 
and  the change in the internal energy of the bubble goes entirely towards raising the internal energy of the ambient medium.
In this case, the total energy available for heating the ICM is simply that due to the change in the potential energy and the
corresponding heating rate (in units of energy per unit time) is
$\vb F_{_{\rm bu}}=(4\pi/3)\rb^{3} \rhoa g_{_{\rm eff}}\vb\;.$
To get the heating rate per unit mass, ${\caH}_{_{\rm heat}}$,
we multiply this expression
by the number density of bubbles, $\nb$, and divide it by the mass density of the ambient
medium per unit volume $\rhoa$ (see equation \ref{eq:rhoadef}).
The result is  \begin{equation}
{\caH}_{_{\rm heat}}=\frac{4\pi}{3}\rb^{3} \nb g_{_{\rm eff}} \vb \; .
\label{eq:hheat}
\end{equation}
It is instructive to relate this heating term to the
physical parameters of the bubbles and their injection rate.
Assume that bubbles are injected within a sphere of radius $\rej$
around the center, at a rate $\dnej$ per unit volume.
Neglecting bubble destruction, the total number  of bubbles within a distance $r\gg \rej$
is $\dnej (4\pi/3) \rej^3 r/\vb$, so that the number density within $r$ is $~\bar \nb(<r)\approx
\dnej \rej^3/(\vb r^2) $. Approximating  $\nb(r)\sim \bnb(r)$
in (\ref{eq:hheat}) and using  (\ref{eq:vbub}) gives   \begin{equation}
{\caH}_{_{\rm heat}}\sim \frac{4\pi}{3} \dnej {\rej^3}  g_{_{\rm eff}} \frac{\rb^3 }{r^2  } \; .
\end{equation}
It is interesting that this does not depend on the drag coefficient,
$\cd$. We can further simplify this expression by noting
the bubble energy at $r$ is $\eb\sim
4\pi \rb^3 p/3/(\gammab-1)$. The final result is
\begin{equation}
{\caH}_{_{\rm heat}}\sim
(\gammab-1) \dnej  g_{_{\rm eff}} \frac{\rej^3}{r^2}  \frac{\eb}{ p } .
\end{equation}
For $r<\rej$ the heating rate is obtained by replacing $\rej$ with $r$ in the last expression.

In the numerical code, the heating is computed  explicitly using (\ref{eq:hheat}).  We do so because apart from the
initial acceleration phase, the bubble's motion should be well described by the ``slow rise'' approximation.  In
the worst case scenario, we are underestimating heating rate in \phaii by a factor of $1+1/\gammab$.  This uncertainty
is not of concern because in
our adopted scenario where the bubbles are initially injected in a highly overpressurized state
($p_{{\rm 1}} \ll \pbi $), the heating of the ICM during \phai dominates over the heating by the rising, expanding bubble.

%%%%%%%%%%%%%%%
%%%%%%%%%%%%
%%%%%%%%

\section{Self-regulating feedback}
\label{sec:self}

We write the energy equation in terms of the ICM entropy $\caSi=\ui/\rhoi^{\gammai-1}$ as 
\begin{equation}
\rhoi^{\gammai-1}\dot {\caSi}={\caH}-\frac{u}{t_{\rm cool}}
\label{eq:sene}
\end{equation}
where $t_{\rm cool} $ is given in Appendix B.
As explained before, two sources contribute to  ${\cal H}$:
$a)$  heating by weak shocks generated during the initial rapid expansion of 
the over-pressurized  bubbles, and $b)$  heating by drag with the ICM as bubbles
rise up by buoyancy. Adiabatic work done by the bubbles 
as they move upward does not change $\caSi$ and therefore does not 
contribute to $\caH$. 
 We 
introduce $\rej$,  the distance from the center within
which bubbles are injected. 
At $r>\rej$, bubbles can be present only as a result 
of flow of bubbles from regions with $r<\rej$. 
Given the injection rate  (number per unit volume per unit time), $\dnej(t,r)$,
of bubbles, we write 
the local heating rate (energy per unit mass of the ICM) via weak shocks
as 
\begin{equation}
\caH_{_{\rm wsh}}=\dnej\frac{\Delta E_1}{\rhoa} \; ,
\label{eq:wsh}
\end{equation}
where $\Delta E_1$ is given by (\ref{eq:Edissp}). 
As a recipe for self-regulated feedback, we assume that bubbles
 are generated only if $\dot \caSi$ is negative.
We write the flux 
of injected bubbles at a point $r\le \rej$ as 
\begin{equation}
\dnej=-\frac{\eta}{\ebi} \frac{\dot \caSi}{\caSi}  \rhoi c^2\quad {\rm for} \quad \dot \caSi <0\; ,
\label{eq:dnej}
\end{equation}
and zero, otherwise. The free parameter  $\eta $ 
represents 
the product of the mass fraction of cold gas that accretes onto the AGNs
and the efficiency of AGNs at transforming the accreted mass into bubbles. 
The cold gas is assumed to be generated at a rate (in units of mass
per unit volume per unit time) of $\rhoi \dot \caSi/\caSi$. This cold
gas is assumed to form by condensation via the development of 
thermal instabilities (e.g. Field 1965).
Finally, $c$ is the speed of light and, as before, $\ebi$ is the 
energy with which a bubble is injected.

{ We can use the above formalism to investigate the role of heating by weak shocks, the dominant
source of heating in our model. Substitute  $\Delta E_1$ and $\dnej$ from (\ref{eq:Edissp})
 and (\ref{eq:dnej}), respectively, into (\ref{eq:wsh}) yields
 \begin{equation}
 \caH=-\uA\frac{\dot \caSi}{\caSi}\; , \quad {\rm where}\quad
 \uA=\cwsh \eta c^2 \left[1-\left(\frac{p}{\pb}\right)^{1-\frac{1}{\gammab}}\right] \; ,
\label{eq:cahua}
\end{equation}
where $p=p(r)$ is the ambient pressure at $r$ and $\pb$ is the internal bubble 
pressure with which the bubbles are injected.
This expression holds for $\dot \caSi<0$. For $\dot \caSi>0$, we
have $\caH=0$. 
Substituting (\ref{eq:cahua}) in (\ref{eq:sene}) and solving for 
  $\dot \caSi$  we 
get,
\begin{equation}
\dot \caSi=-\frac{\caSi}{\tau_{\rm cool}} \; ,
\label{eq:SA}
\end{equation}
where 
\begin{equation}
\tau_{\rm cool}=\left(1+\frac{u_{_{\rm A}}}{u}\right)t_{\rm cool} \; .
\end{equation}
The above equation implies that cooling is substantially suppressed at low temperatures, 
$u\ll u_{_{\rm A}}$, while at higher temperatures, shocks are unable to suppress cooling. 
For an ICM at temperature corresponding 
to velocity  $V$
an efficient suppression of cooling at distance $r$ occurs if
\begin{equation}
\cwsh \eta \sim \left(\frac{V}{c}\right)^2\approx 10^{-5}\left(\frac{V}{10^3\kms}\right)^2\; ,
\end{equation}
where we have assumed $\pb\gg p$.
Note that since $\dnej \propto 1/\ebi$, neither the shock nor the drag heating depend on $\ebi$.}

\section{Numerical model}
\label{sec:numerical}
We develop a numerical  hydrodynamical model to describe the 
evolution of the ICM and the bubbles in a  
fixed dark matter halo.
The density profile in the halo is approximated by the 
NFW profile (e.g. Hayashi \etal 2004) of the form
\begin{equation}
\rho(s,t)=\frac{\delta_{\rm c} \bar \rho(t)}{s(1+c s)^2}\; ,
\end{equation}
where $s=r/R_{\rm v}$ and $R_{\rm v}$ is the virial radius of the halo. 
The parameter $\delta_{\rm c}$ is determined by 
the condition that the mean density within $r=R_{\rm v}$ is $178$ times
the background density, $\bar \rho(t)$. 
We adopt the value $c=5$ for the concentration parameter (e.g. Balogh \etal 2005).

The ambient medium is represented by lagrangian shells
of fixed mass. In each time-step, the equations of motion of the bubble  and the ambient 
fluids are solved in the following order:

\noindent $\bf (i)$ advance the ambient shells according to equations 
(\ref{eq:euler}) and (\ref{eq:adiab})  without radiative cooling or 
dissipative heating.

\noindent $\bf (ii)$ update bubble properties in each shell as a result of the change in 
incurred in the ambient medium (see Appendix A) after the hydrodynamic step (i).

\noindent $\bf (iii)$ estimate dissipative  heating (by weak shocks and drag)  and radiative cooling of ICM and update the corresponding ambient pressure and 
 bubble properties  without moving the ambient shells. 

\noindent  $\bf (iv)$ inject the ambient medium with new bubbles according to the recipe 
 outlined in (\S\ref{sec:self}).
 
The integration of the hydrodynamic equations (\ref{eq:euler}) and (\ref{eq:adiab})   includes artificial viscosity (e.g. Richtmyer \& Morton 1967;  Thoul \& Weinberg 1995; Nusser \& Pointecouteau 2006) in order to deal with 
any abrupt changes in the velocity which may arise if a cooling runaway 
develops.  
The bubble fluid is moved through the shells of the 
ambient medium using a first order flux conserving scheme.
The velocity of this fluid relative to the shells is given by (\ref{eq:vb}).
Finally, a leapfrog time integration scheme is used.

\section{Results}
\label{sec:results}

We adopt  a flat cosmology  with  baryonic  and dark matter
density parameters respectively  $\Omega_{\rm b}=0.044$ and  $\Omega_{\rm m}=0.27$, a cosmological constant corresponding to a
density parameter of $\Omega_{\Lambda}=0.686$, and a Hubble constant of 
$H_0=70\rm km \; s^{-1} Mpc^{-1}$ (e.g. Spergel \etal 2003).

The number of shells describing the ambient medium is $50$ in all of the runs. 
The initial shells are placed at logarithmically spaced distances from 
the center with the nearest shell $r=15\rm kpc$ and the farthest lying at the 
virial radius of the halo.
The initial temperature profile is assumed to be isothermal with 
a temperature,  $T$ (in keV), given by $1.16 \times 10^7 k_{\rm B} T/\mu m_{\rm p}=(2/3) G M_{\rm v}/R_{\rm v}$ where 
$k_{\rm B}$ is the Boltzmann constant, $\mu$ is the mean molecular weight  and 
$m_{\rm p}$ is the mass of the proton. The equation for hydrostatic equilibrium 
implies  a slope of $-9/4$ for the gas density at $r=R_{\rm v}$. 
The gas density at any $r$ ($r<R_{\rm v}$) is solved numerically using the equation of hydrostatic
equilibrium including gas gravity. In solving this equation, 
 the  gas to dark matter mass 
ratio at $r=R_{\rm v}$ is taken to be  the background ratio $\Omega_{\rm b}/\Omega_{\rm m}$.
The mass in each ambient shell is then estimated from the gas density profile. 
Because of the limited number of shells, a constant temperature does not necessarily 
guarantee that the shells are actually in strict hydrostatic equilibrium. Therefore, 
given the mass in each shell, we use  the discrete form of the equations of motion as 
they appear in the numerical hydrodynamical code (e.g. Weinberg \& Thoul 1995) to
obtain a temperature (thermal energy per baryon) in each shell such that
strict equilibrium is achieved at the initial time. 
This yields nearly isothermal temperature profiles with deviations 
that increase with radius.
 
In all runs, the code evolves the shells from redshift $z=2$ until $z=0$.
In some runs with insufficient feedback to counter cooling, the 
hydrodynamic time-step becomes exceedingly small. Whenever this happens, the 
shells colder than $1\; \rm keV$ and denser than $1.5 \rm cm^{-3}$ 
are held fixed and are not moved by the code at any later time.

\begin{figure*}
\centering
\begin{sideways}
\mbox{\psfig{figure=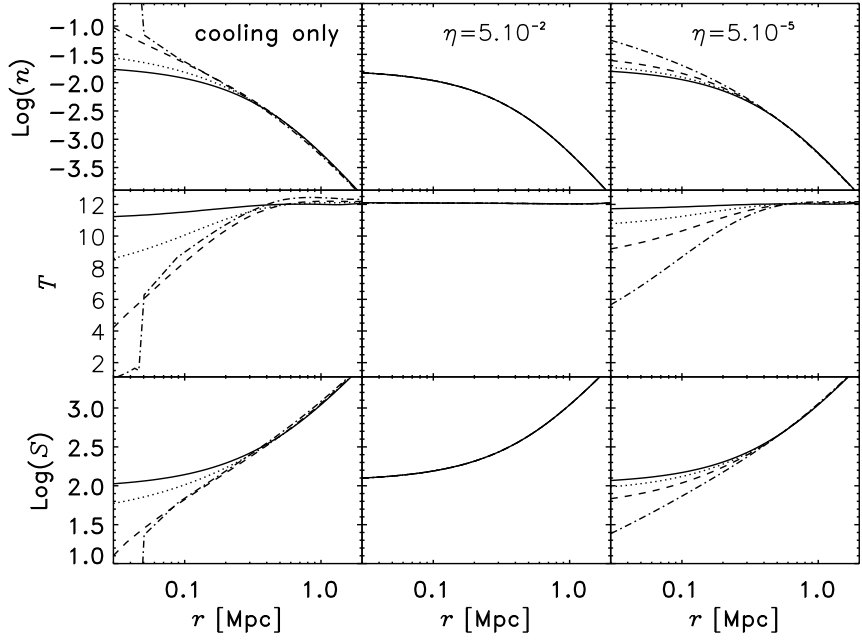,height=4.5in}}
\end{sideways}
\vspace{0.25cm}
\caption{The ICM electron number density (top), temperature (middle) and entropy (bottom) profiles versus   distance from the cluster center (in $\rm Mpc$). 
The left column shows profiles obtained with radiative cooling and no energetic feedback.
In the  middle and right panels feedback is done via bubbles generated 
in the entire cluster with  efficiency parameter $\eta=10^{-2}$ and 
$\eta=10^{-5}$, respectively.
 The solid, dotted, dashed and dash-dotted curves correspond to 
 redshifts $z=1.6$, $1$, $0.5$, and $0$, respectively. The results correspond to an
 NFW dark halo profile with a virial radius of $R_{\rm v}=4\; \rm Mpc$ and 
  a concentration parameter of $c=5$. The number density,  temperature, and entropy are given  in $\rm cm^{-3}$, $\rm keV$,
and $\rm keV cm^2$, respectively.}
\label{fig:fig4mpc}
\end{figure*}

\begin{figure*}
\centering
\begin{sideways}
\mbox{\psfig{figure=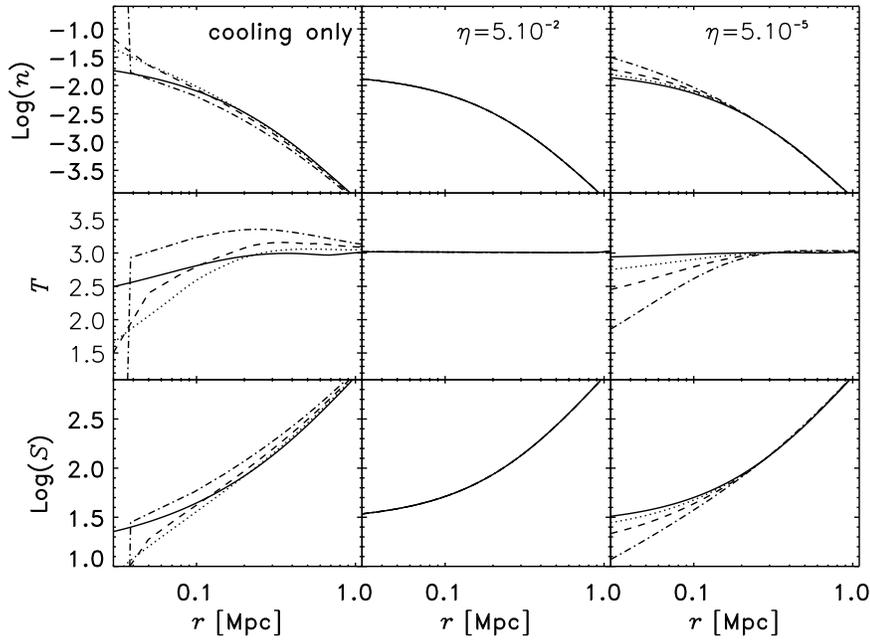,height=4.5in}}
\end{sideways}
\vspace{0.25cm}
\caption{The same as the previous figure but for a dark halo with a virial radius of
$R_{\rm v}=2\rm Mpv$.}
\label{fig:fig2mpc}
\end{figure*}

Results from simulations for clusters with virial radii of $\rv=4\; \rm  Mpc$ and 
$\rv=2\; \rm Mpc$
are shown in Fig.~\ref{fig:fig4mpc} and Fig.~\ref{fig:fig2mpc}, respectively.  
In the column to  the left we show results which include radiative cooling and no 
heating.  For both values of $R_{\rm v}$, effects of cooling become 
significant already by $z=1$. At $z<0.5$ a cooling runaway develops 
in the inner regions: cooling becomes increasingly more rapid 
due to the enhanced density and lower temperature. 

The effect of heating by bubbles, according to our feedback 
recipe described in \S\ref{sec:self}, is explored in the 
middle and right panels for two values of the efficiency parameter $\eta$.
To model the effect of multiple AGN activity, bubbles are allowed to be injected 
anywhere in the ICM (i.e. $\rej=R_{_{\rm v}}$).
As noted before the heating rate of the ICM does not depend on 
 the initial  bubble energy, $\ebi$.
All results in these two columns are for $\cd=1$ and $\cwsh=1$, and an initial bubble pressure of $10^3$ times the central ICM pressure. 
Cooling flows for the two different clusters are completely  suppressed
for $\eta$ as low as 0.01, while a value $\eta=10^{-5}$ gives moderate suppression of cooling as is required by the observations.  

\section{Summary and discussion}
\label{sec:sum}

We  propose that modest radio outbursts from many cluster galaxies may help to quench 
cluster cooling flows.  We appeal to typical massive elliptical galaxies which are thought 
to have radio outbursts with a duty cycle of $\sim 10^8$ yr per Hubble time (Best \etal 2006) 
and are concentrated in the cluster core (within the central $~200 \; \rm kpc$ or of order 
twice the extent of the cooling region). Weak shocks generated during the initial bubble 
expansion and viscous/turbulent drag on buoyantly rising bubbles are the principal dissipative 
heating processes.  Indirect evidence for multiple heating sources in the form of AGN comes 
from the  frequent presence of x-ray ``ghost'' cavities that survive long after any associated radio 
lobes have decayed, and from amorphous radio halos that further suggest an ICM 
percolated with a distribution of bubbles whose contrast in the X-ray has been diminished 
due to adiabatic expansion (cf. Heinz \& Churazov 2005). Our scenario is in agreement with 
recent observational evidence for  the existence of AGN activity in about 5$\%$ of cluster 
galaxies at low redshift (Martini \etal 2006).  Cosmological evolution of the AGN inevitably 
increases the feedback at earlier epochs ($z\ltsim 2$).  
Our scenario alleviates some of the problems that may arise if  energy feedback is generated by
a single central source (AGN) harboured in the central cD galaxy. The  needed energy transport 
over the entire cooling flow region is not easy to provide if the  energy source is associated 
only with the central AGN (e.g. Mathews, Faltenbacher \& Brighenti 2005).  
One approach is to ensure that mechanical disturbances are able to propagate large distances before 
they are damped by viscosity (e.g. Ruszkowski, Br\"uggen \&  Begelman 2004; Reynolds \etal 2005). 
This approach has two potential shortcomings.  Firstly, the heating rate per baryon of dissipating 
waves should decay with distance from the central source as $r^{-3}$. This  is too steep a 
gradient for the heating to affect the ICM over the entire region that is susceptible to cooling.
In fact, Roychowdhury \etal (2004) find that the heating leads to a convectively unstable core 
profile that would potentially destroy any pre-existing metallicity gradient, in conflict with 
the x-ray observations,  and to remedy this,  Roychowdhury et al.  (2005) invoke relatively high 
values of thermal conduction to effect rapid transport of the energy out of the center.  
Secondly, the viscosity has to be finely tuned: too high a value results in dissipation much to closely
concentrated near the central source, and too low a value means the heating is inefficient. Since the 
relevant viscosity coefficient scales as $T^{5/2}$ (Spitzer 1962), it is difficult to envison that 
a viscous solution that relies on fine-tuning prevails in all clusters.  Moreover, the ICM is weakly
magnetized, and the nature of transport processes, like thermal conduction and viscosity, in such media
is not well understood.

{ 
The   model presented here  heats the ICM  through shocks generated 
 during the initial inflation of the bubbles, and  drag forces acting on the rising bubbles.
However, the latter relies on some viscosity to dissipate heat but it is not an important source of heating and with respect to the former, we do not
need to fine-tune the viscosity since the energy distribution is effected by multiple AGNs hosted in 
a spatially extended distribution of galaxies.  Moreover, it may be possible to avoid the issue 
altogether.   Heinz \& Churazov (2005) have recently argued that in a
bubble-filled medium of the kind proposed here, the bubbles themselves act as catalysts to convert 
shock and waves into heat.  Additionally,  the interaction between the bubbles and the strong shocks 
will both broaden and distort the shock fronts, causing the features to appear weaker in X-ray observations 
and resulting in mistaken assumptions that cavity inflation is a gentle process  and that the ICM is 
not subject to strong shocks.  Both of these features further contribute to making our model highly 
viable.}

An additional principal feature of our model is the incorporation of self-regulation between heating and cooling. This should inevitably lead to  a correlation between the mechanical heating luminosity and the cooling rate. The former is measured by the properties of x-ray cavities and the latter by the cluster x-ray luminosity. Evidence for such a correlation is given by 
 B\^irzan \etal (2004).
 
 In fact, the estimated ratio of mechanical to x-ray luminosity in this latter paper falls short of that needed by Croton \etal (2005) in order to quench cooling flows on group and cluster scales by a factor of 3-10 (Best \etal 2006). However, the inferred mechanical luminosities are based on 
 cavity $pV$ estimates. In our model, a primary source of  energy transfer 
 to the ICM is via weak shocks generated by the expansion of
the initially over-pressurised bubble. The
initial bubble energy 
 is related to the cavity $pV$ by equation (\ref{eq:eb0}), and the resulting injected mechanical energy may exceed $pV$ by a factor of up to 10.
 Note that this equation neglects the work done by the AGN jet to 
 place the initial seed bubble in the ICM, which can be dissipative and 
 contribute to heating the ICM if the seed bubble is ejected with supersonic 
 speeds.  
 Furthermore, these estimates neglects the energy content in waves 
 detected in a few clusters (Fabian \etal 2003; Nulsen \etal 2005, McNamara \etal 2005). Our model consequently helps in resolving the apparent discrepancy in the estimated feedback required to halt cooling flows even in massive clusters. 
 
An alternative heating source involves cosmic rays in the ICM cooling core, injected by a central AGN (Rephaeli and Silk 1995). The present model
provides a distributed source of cosmic rays whose observable signatures could also include a Sunyaev-Zeldovich signal (Pfrommer \& Ennslin 2004) as well as an additional hadronic heating source (Colafrancesco,
Dar \& De R\'ujula 2004) that might lead to a  detectable gamma ray flux.
More exotic heating sources  invoke collisions between superheavy 
dark matter particles and protons (Qin \& Wu 2001, Chuzhoy \& Nusser 2004)
as well as by the products of dark matter neutralino annihilations (Colafrancesco 2004).

\section{Acknowledgments}
We acknowledge stimulating discussions with Scott Kay, Garret Cotter and Yehuda
Hoffman.  We would also like to thank the anonymous referee for most
helpful comments that contributed to the improved clarity of the paper.
This work was initiated while A.B. and A.N. were visiting the
Astrophysics Department at the University of Oxford.   They would like
to expressed their gratitude for the  hospitality and generous support 
shown to them.  A.B. would also like to acknowledge support from the Leverhulme 
Trust (UK) in the form of the Leverhulme Visiting Professorship.

\protect

\appendix
\label{app:not}

\section{Thermodynamics of the bubbles and ICM}
\subsection{Numerical description of  bubbles in ambient medium}
For simplicity the mass of bubbles is neglected.

As bubbles are transported in the ICM in an infinitesimal time step,
the ICM in a shell
is either compressed or diluted depending on whether or not  number of
bubbles in the shell has increased or decreased. 
This is a purely adiabatic process that conserved  
the entropies inside a bubble and 
in the ICM.

Consider a shell with pressure $p_1$ containing bubbles each of radius $R_{\rm b1}$. Assuming that $\delta N$ additional bububles enter the shell, we would like to estimate the ambient pressure, $p$, the radius, $R_{\rm b}$, and the filling factor, $F$, after the bubbles enter the shell. 
For concreteness let  $\delta N>0$, i.e., there is a  net increase in the 
number of bubbles in the shell.  
Before these bubbles enter the shell, the total material  
in the shell occupies a volume $V_1=4\pi(r_{i+1}^2-r_i^3)/3$ of which 
$V_{b1}=F_1 V_1$ is taken by the bubbles and $V_{I1}=(1-F)V_1$ by the ICM.

After $\delta N $ bubbles enter the shell, 
the same original 
material occupies a volume $V_1-\delta V$ where $\delta V=\delta N
4\pi \rb^3/3$ with  $\rb$ being  the new radius of a bubble in the shell.
The entropies in the bubbles and the ICM of the original 
are conserved so that $p V_I^\gammai=p_1 V_{I1}^\gammai$ and 
$p V_b^{\gammab}=p_1 {V_{b1}}^{\gammab}$ where $V_I$ and $V_b$ are the volumes
occupied by the ICM and bubbles of the original material after the $\delta N$ bubbles entered the shell. Using these entropy conservation relations in 
$V_1-\delta V=V_I+V_b$ gives
\begin{equation}
1-\frac{\delta V}{V_1}=
F_1
\left(\frac{p_1}{p}\right)^{\frac{1}{\gammab}}+\left(1-F_1\right)
\left(\frac{p_1}{p}\right)^{\frac{1}{\gammai}} \; .
\label{eq:newbubb1:adiab}
\end{equation}
Entropy conservation inside a bubble gives 
 $\rb^3=\rb_1^3(p_1/p)^{1/\gammab}$ which after substituting in 
  $\delta V=\delta N  4\pi \rb^3/3$ yields, 
\begin{equation}
1=\left(F_1+\delta N \frac{4\pi}{3}\rb_1^3\right)
\left(\frac{p_1}{p}\right)^{\frac{1}{\gammab}}+\left(1-F_1\right)
\left(\frac{p_1}{p}\right)^{\frac{1}{\gammai}}\; .
\end{equation}
The bubble radius and their filling factor after the 
change can easily be computed from the adiabatic conditions once
the pressure, $p$, is known.

\subsubsection{Effect of heating and cooling on the ICM and bubbles}
As mentioned previously, the ICM is meant to describe the inter-cluster plasma 
lying outside the bubbles. 
As an infinitesimal amount of energy is removed from the ICM by cooling 
or added, we must describe the reaction of 
the ICM and bubbles to such infinitesimal change of the energy.
Let an  energy  $\delta E$ be added as heat to the ICM 
in the shell $i+1/2$.  For concreteness we assume $\delta E>0$, but the same reasoning applies to cooling $\delta E<0$.
In the calculation we assume that the that the system 
reacts in the following two step sequence. First the 
pressure of the ICM is changed to $p_1+(\gammai-1)\delta E/V_0$
where $V_0=4\pi (r_{i+1}^3-r_i^3)/3$ before the bubbles
are compressed. Next, the under-pressured bubbles are 
compressed by the ICM.
The expansion of the ICM and the compression of the bubbles
in the second step are both assumed to be adiabatic processes. 
The following adiabatic conditions hold in the 
second step.
\begin{equation}
p V_I^\gammai=\left[p_1+(\gammai-1)\frac{\delta E}{V_{I1}}\right]
V_{I1}^\gammai
\quad {\rm and }\quad p V_b^{\gammab} =p_1 V_{b1}^{\gammab}\; 
\end{equation}
where $p$, $V_{I}$, and $V_b$ are the pressure, the volume of 
the ICM, and the  volume of bubbles at the end of the second step. 
These conditions give the following equation for $p$
\begin{equation}
1=F_1\left(\frac{p_1}{p}\right)^{\frac{1}{\gammab}}
+(1-F_1)
\left[\frac{p_1}{p}+(\gammai-1)\frac{\delta E }{F_1  V_0 p}\right]^{\frac{1}{\gammai}} \; .
\end{equation}

\section{The cooling function and the X-ray emissivity}
\label{app:cooling}
 We approximate the cooling time by (Balogh, Babul \& Patton 1999)
 \begin{equation}
 t_{\rm cool}=\frac{B_1\mu m_{\rm H} T_{_{\rm I}}(r)^{1/2}}
 {\rhoi\left[1+B_2f_{\rm m}/T_{_{\rm I}}(r)\right]}\;  ,
 \label{eq:tcool}
\end{equation}
where, $\mu$ is the mean molecular weight,  $B_1=3.88\times 10^{11}\rm s K^{-1/2}cm^{-3}$,
$B_2=5\times 10^7\rm K$ and $f_{\rm m}$ is a metallicity-dependent
constant  that is 1 for solar metallicity and $0.03 $ for zero metallicity. 

The volume X-ray emissivity of the ambient medium
is
\begin{equation}
\epsilon=\frac{3}{2}\frac{(1-F)\rhoi}{\mu m_{\rm H}}
\frac{k T_{\rm I}}{t_{\rm cool}}
\end{equation}

\end{document}